\pdfoutput=1
\documentclass[prb,preprint,nofootinbib]{revtex4-2}

\usepackage[colorlinks, allcolors=blue]{hyperref}
\usepackage{natbib}
\usepackage{amsmath}  
\usepackage{amsfonts} 
\usepackage{graphicx} 
\usepackage{mathrsfs}
\usepackage{soul,xcolor}

\begin{document}
\setstcolor{red}

\title{Unveiling the interdisciplinary character of negative pressure}

\author {Francisco F. Barbosa}
\affiliation{São Paulo State University (UNESP), Institute of Geosciences and Exact Sciences, Rio Claro - SP, Brazil}
\author {Lucas Squillante}
\affiliation{São Paulo State University (UNESP), Institute of Geosciences and Exact Sciences, Rio Claro - SP, Brazil}
\author{Luciano Ricco}
\affiliation{Science Institute, University of Iceland, Dunhagi-3, IS-107 Reykjavik, Iceland}
\author {Roberto E. Lagos-Monaco}
\affiliation{São Paulo State University (UNESP), Institute of Geosciences and Exact Sciences, Rio Claro - SP, Brazil}
\author{Antonio C. Seridonio}
\affiliation{São Paulo State University (Unesp), Department of Physics and Chemistry, Ilha Solteira - SP, Brazil}
\author {Mariano de Souza}
\email{mariano.souza@unesp.br}
\affiliation{São Paulo State University (UNESP), Institute of Geosciences and Exact Sciences, Rio Claro - SP, Brazil}

%


\author{\textmd{Submitted on November 06$^{\textmd{th}}$, 2023}}

\begin{abstract}
We explore the concept of negative pressure and its relevance in a variety of physical contexts: the expansion of the universe, mixture theory, cavitation, and the capillary effect in plants. Using thermodynamic arguments, we discuss the intricate connection between negative pressure and negative thermal expansion. We highlight the fact that metastable states and competing phases are often associated with the emergence of negative pressure. We also propose a new link between the effective Gr\"uneisen parameter and nucleation theory.
\end{abstract}

\maketitle

\clearpage

\section{Introduction}

Exotic physical phenomena arise under extraordinary conditions, such as very strong high-magnetic fields, extreme pressures, or at low-temperatures. Here, we focus on the unusual situation of a system with \emph{negative} pressure. The definition of pressure $p$ is familiar from introductory physics, $p=F/A$, where  $F$ is the modulus of the force component acting on area $A$. Throughout the manuscript we have adopted the convention that lower-case letters are used for intrinsic physical quantities. In thermodynamics, $p$ is defined as $p=-\left(\frac{\partial \mathcal{F}}{\partial v}\right)_T$, where $\mathcal{F}$ is the Helmholtz free energy, $T$ temperature, and $v$ the volume.

\begin{figure}[!h]
\centering
\includegraphics[width=0.86\textwidth]{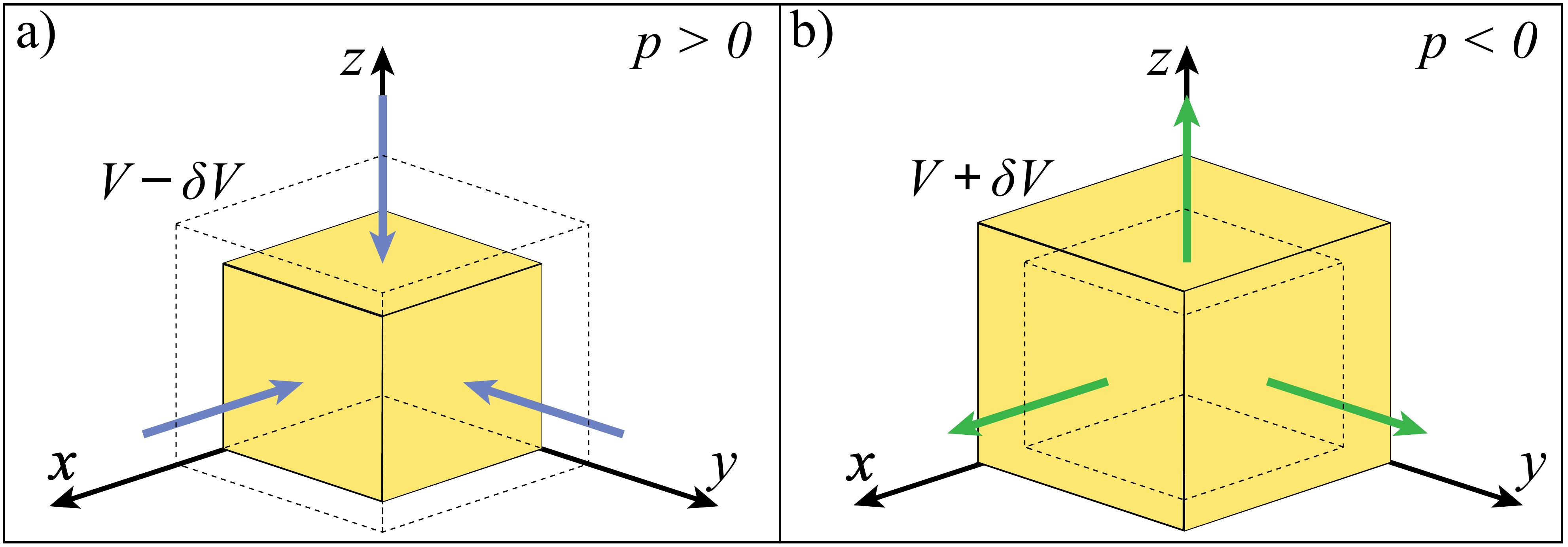}
\caption{\footnotesize Schematic representation of a solid cube under pressure $p$ which is a) positive and b) negative. The dashed lines represent the original volume $V$ in the absence of $p$. In a), the blue arrows indicate the compressive force acting on a cube decreasing its volume $V$ by $\delta V$, while in b) the green arrows represent the tensile force increasing $V$ by $\delta V$. Such a representation is applicable when metastability sets in, giving rise to the emergence of negative pressure, which is linked with tensile stress (See Sec.\,\ref{tensilestress}).}
\label{Fig-0}
\end{figure}

Negative pressure can appear in biology \cite{nobel}, cosmology \cite{barbara}, and solid state physics \cite{baidakov}. Negative pressures can be attained experimentally, for instance, via application of tensile stress by a piezoelectric actuator, increasing the lattice parameters of the system \cite{ikeda}. A simple way to think about negative pressure is by associating it with an outward force applied over a surface area (this is tensile stress in a solid), whereas an inward force is associated with a positive pressure (or compressive stress in a solid), see Fig.\,\ref{Fig-0}. Putting it another way, when negative pressure emerges, the force is in the opposite direction from what we usually would expect.

Landau and Lifshitz's textbook \cite{Landau} links negative pressures with metastable states, using the relationship $\frac{p}{T} = \left(\frac{\partial S}{\partial v}\right)_U$, where $S$ is the entropy and $U$ the internal energy. Note that, since $T > 0$, for $\left(\frac{\partial S}{\partial v}\right)_U > 0$ $\Rightarrow$ $p>0$, meaning that $S$ increases when the system expands, whereas for $\left(\frac{\partial S}{\partial v}\right)_U < 0$ $\Rightarrow$ $p<0$, so that $S$ increases upon decreasing $v$, which is, at first glance, counter-intuitive because usually decreasing $v$ implies lower $S$ due to less accessible states, cf.\,Fig.\,2.9 of Ref.\,\cite{thermalphysics}. Landau and Lifshitz elucidate that the genesis of negative pressures lies in the inherent metastability associated with two distinct competing phases/configurations.
\begin{figure}[!h]
\centering
\includegraphics[width=0.65\columnwidth]{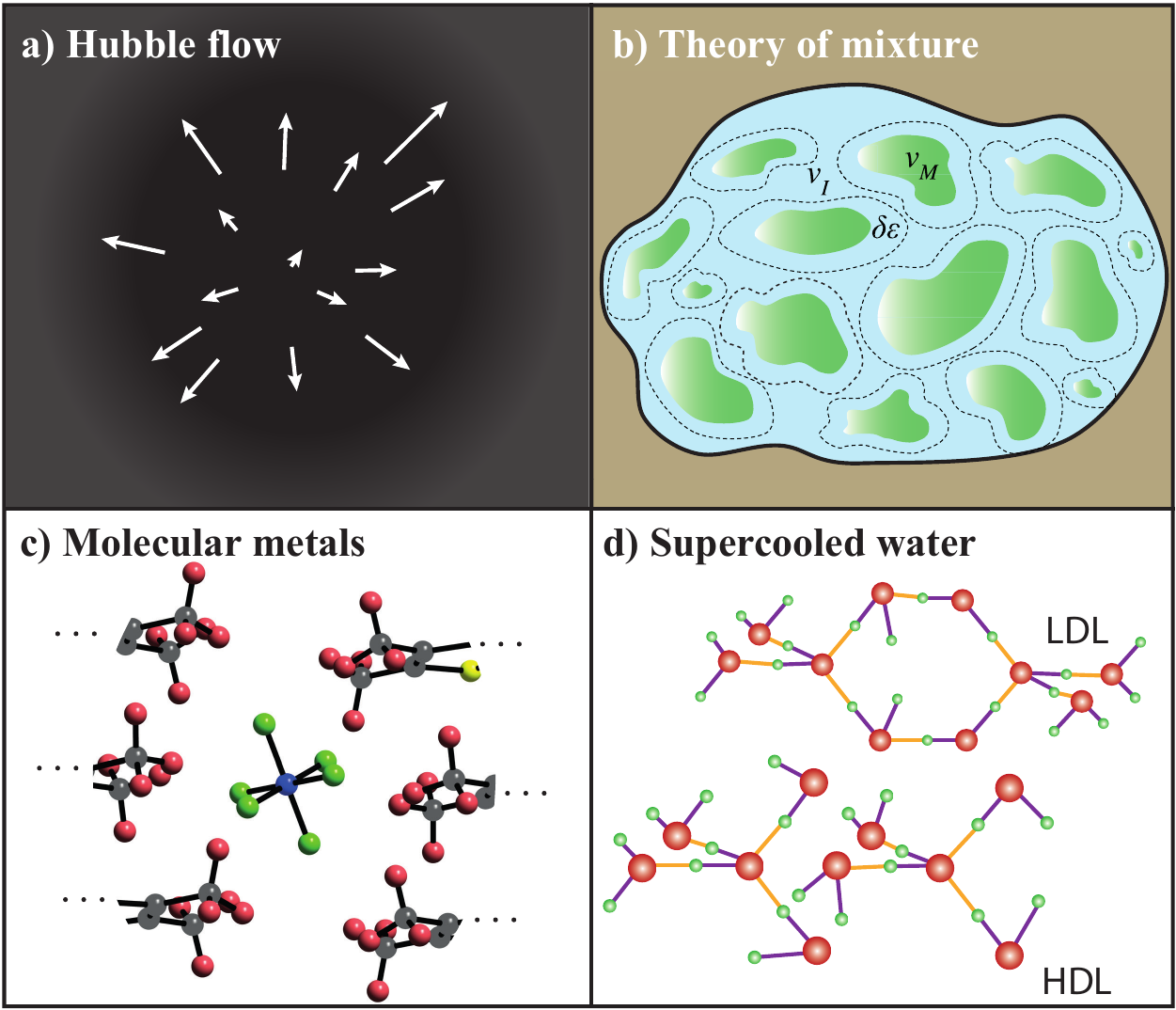}
\caption{\footnotesize (Color online) Schematic representation of various physical scenarios in which negative pressures occur. a) The accelerated expansion of the universe is explained by considering negative pressures associated with dark energy \cite{barbara}. The white arrows represent recession velocities of galaxies associated with this expansion. b) A binary mixture in which an insulating matrix with volume $v_I$ incorporates metallic puddles with volume $v_M$, and $\delta\varepsilon$ is the mean-field type interaction between neighboring puddles \cite{jap}. c) Monovalent center-symmetric counter-anion trapped in the cavity formed by the methyl-end groups of the TMTTF molecule \cite{prbrioclaro}, where the gray spheres represent C, green spheres F, and the blue spheres Sb, P, or As \cite{prbrioclaro}. The red spheres represent the H atoms, which can be replaced by D emulating negative pressures, see discussions in Ref.\,\cite{pouget}. d) The supercooled phase of water in which high-density liquid (HDL) coexists with low-density liquid (LDL). The intramolecular bonds are indicated in orange color, while the hydrogen bonds in purple. Figure based on Ref.\,\cite{gallo}. In panels b) and d), metastability associated with the coexistence between the two phases/configurations are associated with negative pressures. More details in the main text.}
\label{Fig-1}
\end{figure}
In this scenario, some portion of the diluted phase \emph{detaches} from the vessel's wall, giving rise to cavities in the form of bubbles/droplets. This leads to a spontaneous contraction of the system and the increase of its entropy \cite{Landau}. In other words, droplets/bubbles of the emerging phase are formed within the initially dominant phase. This can happen in the phase coexistence region for a substance on the verge of a first-order phase transition \cite{sethna}. A classic example of this process is cavitation, in which vapor bubbles are formed within the liquid phase below the vapor pressure value \cite{cavitation}.

Because negative pressure appears in many physical contexts across scientific disciplines, it is useful to collect several such applications together in one place; this is one of the aims of the present work. We discuss the cases of the accelerated expansion of the universe, binary mixture theory, molecular metals and chemical pressure, cavitation processes, as well as the capillary effect in plants (see Fig.\,\ref{Fig-1}). We also unveil the key role played by metastability and phase coexistence in the emergence of negative pressure and we explore the link between negative thermal expansion, tensile stress, and negative pressure.

\section{The accelerated expansion of the universe}

It is currently well-known that the expansion of the universe is accelerating \cite{accelerated}. This acceleration is currently understood in the context of the $\Lambda$-CDM model of cosmology. A well-known textbook on the subject is Barbara Ryden's Introduction to Cosmology \cite{barbara}, which is summarized in this section. The dynamics of the expansion of the universe is related to the matter/energy content of the universe through the Einstein field equations \cite{Einstein,barbara}. The Friedmann equations result from the Einstein equations after including assumptions of the homogeneity, isotropy, and a matter/energy content which can be treated as one or more perfect fluids. An accelerating expansion can result if the universe contains a fluid with negative pressure, called dark energy \cite{barbara}. The fact that negative pressure accounts for the accelerated expansion can be seen from the second Friedmann equation, namely \cite{friedmann}:
\begin{equation}
\frac{\ddot{a}}{a} = -\frac{4\pi G}{3c^{2}}(\rho +3p),
\label{friedmann}
\end{equation}
where $a$ is the universe's scale factor, $G$ the universal gravitational constant, $c$ the speed of light, and $\rho$ the energy density\footnote{Note that $\rho$ is the energy density, not the mass density; in SI units [$\rho$] = J/m$^3$ = kg/m$\cdot$s$^2$, so that both $p$ and $\rho$ have the same dimensions.}.
Accelerated expansion of the universe implies that $\ddot{a} > 0$. The equation of state for a perfect fluid is given by $p = \omega \rho$, where $\omega$ is a parameter which depends on the fluid under consideration: for radiation $\omega$ = 1/3, for matter $\omega \rightarrow 0$, and for a cosmological constant $\omega = -1$ \cite{barbara,gruneisencosmology}. Considering the picture of a perfect fluid, the total pressure $p_{tot}$ and energy density $\rho_{tot}$ of the universe is given by the sum of its constituents, namely, radiation (r), matter (m), and dark energy (DE) \cite{barbara}. We consider dark energy as a fluid with $\omega = -1$. Hence, Eq.\,\ref{friedmann} becomes:
\begin{equation}
\frac{\ddot{a}}{a} = -\frac{4\pi G}{3c^{2}}(\rho_{tot} +3p_{tot}) = -\frac{4\pi G}{3c^{2}}[(\rho_{m}+\rho_{r}+\rho_{DE}) +3(p_{m}+p_{r}+p_{DE})],
\end{equation}
so that upon replacing the corresponding $\omega$ values, we obtain:
\begin{equation}
\frac{\ddot{a}}{a} = -\frac{4\pi G}{3c^{2}}(\rho_{m} + 2\rho_{r} -2\rho_{DE}).
\end{equation}
As the universe expands and the scale factor increases, matter and radiation become diluted so that the contributions of both $\rho_{r}$ and $\rho_{m}$ decrease with time. At late times, their contributions can be neglected \cite{barbara}. Thus, we have:
\begin{equation}
\frac{\ddot{a}}{a} \simeq -\frac{4\pi G}{3c^{2}}(-2\rho_{DE}) \simeq  \frac{8\pi G}{3c^{2}}\rho_{DE}.
\end{equation}
Note that to explain observational data associated with an accelerated expansion of the universe \cite{accelerated}, a minus sign in the term $-2\rho_{DE}$ had to be considered. More specifically, a dominant fluid with $\omega < -1/3$ implies $\ddot{a} > 0$. Since $p = \omega\rho$, (and assuming a positive energy density), it becomes clear that the concept of negative pressure in cosmology emerges to account for the accelerated expansion of the universe.

\section{Negative pressures in binary mixtures}

In binary mixtures, negative pressure in the metastable regime is associated with the emergence of a new phase. In this context, we discuss nucleation and spinodal decomposition, as well as the cavitation process.\newline

\subsection{Binary mixture theory and metastable states}
In this subsection we demonstrate that negative pressure can be directly linked with metastability. The following discussions are partially based on Ref.\,\cite{atkins} by Atkins, de Paula, and Keeler. In a metastable binary mixture, the concept of negative pressure
\begin{figure}[!htb]
\centering
\includegraphics[width=\columnwidth]{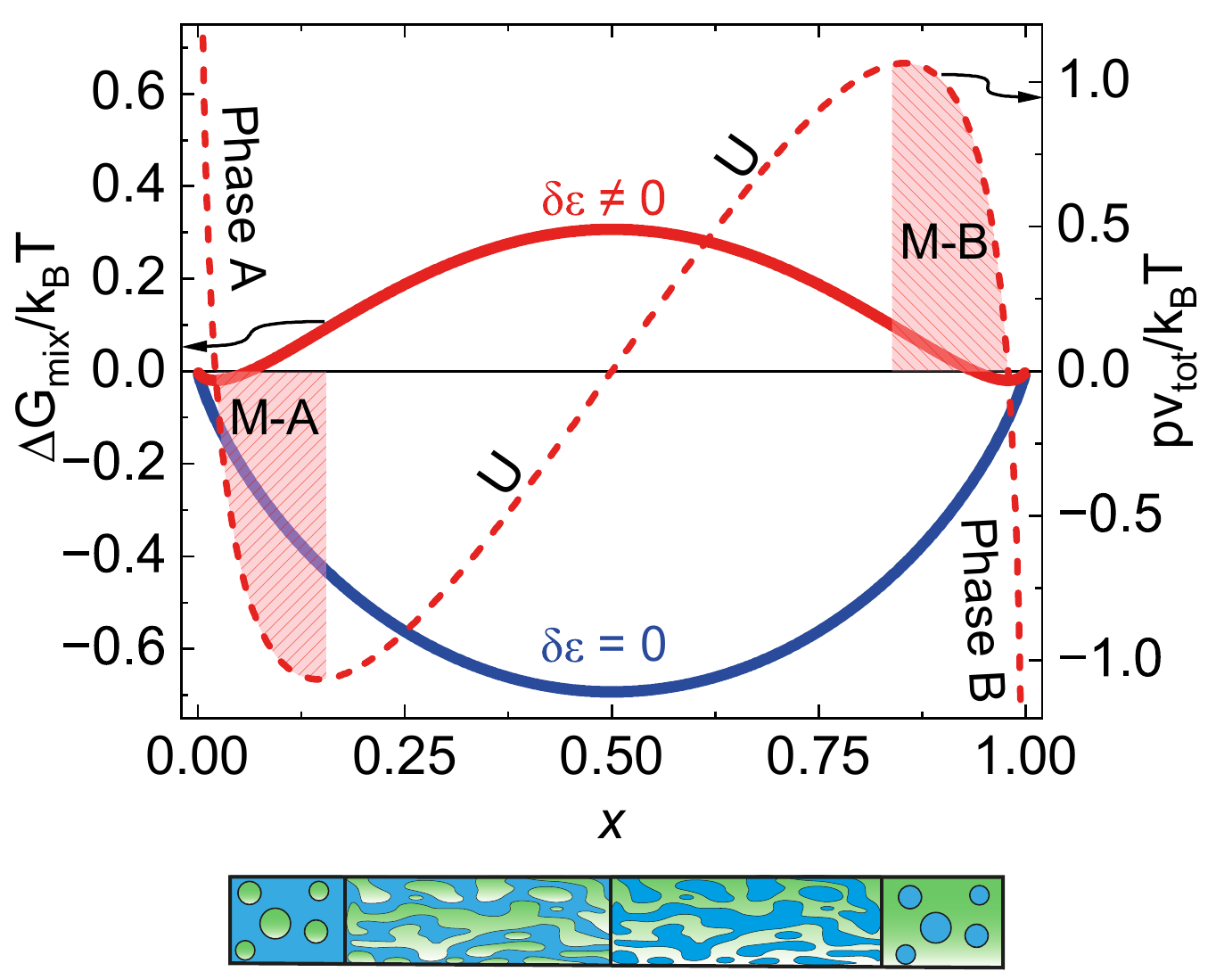}
\caption{\footnotesize (Color online) Left scale: Gibbs free energy of mixture $\Delta G_{mix}$ versus concentration $x$ for $\delta\varepsilon = 0$ (blue solid line) and $\delta\varepsilon/(k_BT) = 3.5$ (red solid line). For $\delta\varepsilon = 0$ no phase separation takes place \cite{atkins}. When $\delta\varepsilon = 3.5k_BT$, the two minima of $\Delta G_{mix}$  delineate the metastable and unstable regimes \cite{Rubinstein}. Right scale: pressure $p$ versus concentration $x$ (dotted red line) normalized by a total volume $v_{tot}$ and thermal energy $k_BT$ considering $\delta\varepsilon = 3.5 k_BT$, showing both metastable phases A and B, respectively, M-A and M-B \cite{sethna}. The stable phases A and B are depicted for, respectively, low and high values of $x$, as well as the unstable regime U. A schematic representation of the corresponding nucleation and spinodal decomposition regimes is also depicted (lower panel). Adapted from Ref.\,\cite{jianguo}.}
\label{Fig-2}
\end{figure}
is governed by the so-called binodal line \cite{atkins}. The Gibbs free energy of mixing per mole, $\Delta G_{mix}$, shown in Fig.\,\ref{Fig-2}, is:
\begin{equation}
\Delta G_{mix} = [x(1-x)\delta \varepsilon] + k_{B}T[x\ln(x) + (1-x)\ln(1-x)],
\label{freeenergy}	
\end{equation}
where $x$ and $(1-x)$ are the concentrations of the constituents (namely $v = xv_{tot}$, where $v$ is the volume of phase A and $v_{tot}$ is the fixed total volume), $k_{B}$ is Boltzmann's constant, and $\delta\varepsilon$ is the mean-field interaction energy. The first term in Eq.\,\ref{freeenergy} accounts for the excess enthalpy, which is zero for an ideal mixture ($\delta\varepsilon = 0$); the second term is due to the entropy of mixing \cite{atkins}. Depending on the magnitude of $\delta\varepsilon$, phase separation can be favoured \cite{Rubinstein}, (see Fig.\,\ref{Fig-2}). By computing $p = -(\partial\Delta G_{mix}/\partial v)_T$, the behavior of $p$ versus $x$ can be obtained. We stress that Eq.\,\ref{freeenergy} describes an incompressible fluid, so that the volume variation in question is related to the volume variation of one of the mixture constituents, or in other words,
\begin{equation}
p = -\left(\frac{\partial \Delta G_{mix}}{\partial v}\right)_T = -\frac{1}{v_{tot}}\left(\frac{\partial \Delta G_{mix}}{\partial x}\right)_T.
\end{equation}
For compressible fluids, the volume dependence of $\Delta G_{mix}$ has to be taken into account \cite{rowlinson}. Regarding Fig.\,\ref{Fig-2}, a single-phase A is established for low values of $x$, which is associated with positive values of pressure. As $x$ is increased, a minimum of $\Delta G_{mix}$ is achieved marking onset of phase separation, so that droplets of a phase B begin to nucleate into phase A. Pressure then becomes negative because of the inherent character of the bubbles of phase B emerging in phase A, as a consequence of the establishment of a metastable phase M-A. As the fraction of phase B is further increased, an unstable regime U is reached in which spinodal decomposition of stripes of phase B compete with phase A. The pressure profile for the stable, metastable, and unstable phases is also shown. Note that the metastable phase B is the one with positive pressure only because, one has to choose which phase is defined to be $x = 0$. The same holds true for the negative $p$ associated with phase B for large values of $x$, i.e., it is just a direct implication of the choice of definition of $x$ \cite{atkins}. Hence, it becomes evident that the positive pressure of one phase is the negative pressure of the other for binary mixtures. In a mixture, negative pressure is associated with phase separation, in which bubbles or droplets appear in a diluted phase \cite{Landau}. Note that (as shown in the lower panel of Fig.\,\ref{Fig-2}) upon reaching the unstable regime, the bubbles merge together, and spinodal decomposition, also called coarsening, takes place following the Cahn-Hilliard equation \cite{cahnhilliard}. Stripes of each phase are formed in this region, demonstrating a mechanism distinct from nucleation.

\subsection{Bubbles embedded in a dominant phase - cavitation}\label{cavitation}
In this subsection we discuss the occurrence of negative pressure in the cavitation process. Cavitation begins with bubble nucleation, which can be described with classical nucleation theory discussed in Ref.\,\cite{sethna} and summarized in the following. The free energy $G_{bubble}$ of a bubble with radius $R$ is given by \cite{sethna,cavitation}:
\begin{equation}
	G_{bubble} (R) = \frac{4}{3}\pi R^{3}(p_l-p_v) + 4\pi R^{2}\sigma,
\label{cavitation}
\end{equation}
where $p_l$ and $p_v$ are, respectively, the liquid and vapor pressures at constant chemical potential and $\sigma$ is the surface tension. More specifically, $p_v$ is the pressure at which the vapor and the liquid share the same chemical potential (for a given temperature)\footnote{Eq. (7) could be rewritten in terms of the value for the vapor temperature for a particular pressure [11].}. Hence, the liquid phase becomes metastable when $p_l$ reaches the value of $p_v$ for a given $T$. Following Ref.\,\cite{sethna}, $G_{bubble} (R)$ has a maximum for a critical radius $R_c$ above which bubble formation becomes energetically favorable, so that an energy barrier $E_b$ can be inferred in terms of $R_c$. In other words, the liquid remains liquid until thermal fluctuations ``pay'' $E_b$ creating thus a gas droplet with a radius $R_c$ \cite{sethna}. At the critical radius, $\partial G_{bubble} (R)/\partial R$ = 0, which yields $R_c$ = 2$\sigma$/($p_v - p_l$). Inserting this expression into Eq.\,\ref{cavitation}, yields \cite{cavitation}:
\begin{equation}
	E_b=\frac{16\pi}{3}\frac{\sigma^3}{(p_v-p_l)^2}.
\label{bindingenergy}
\end{equation}
\begin{figure}[t]
\centering
\includegraphics[width=0.7\columnwidth]{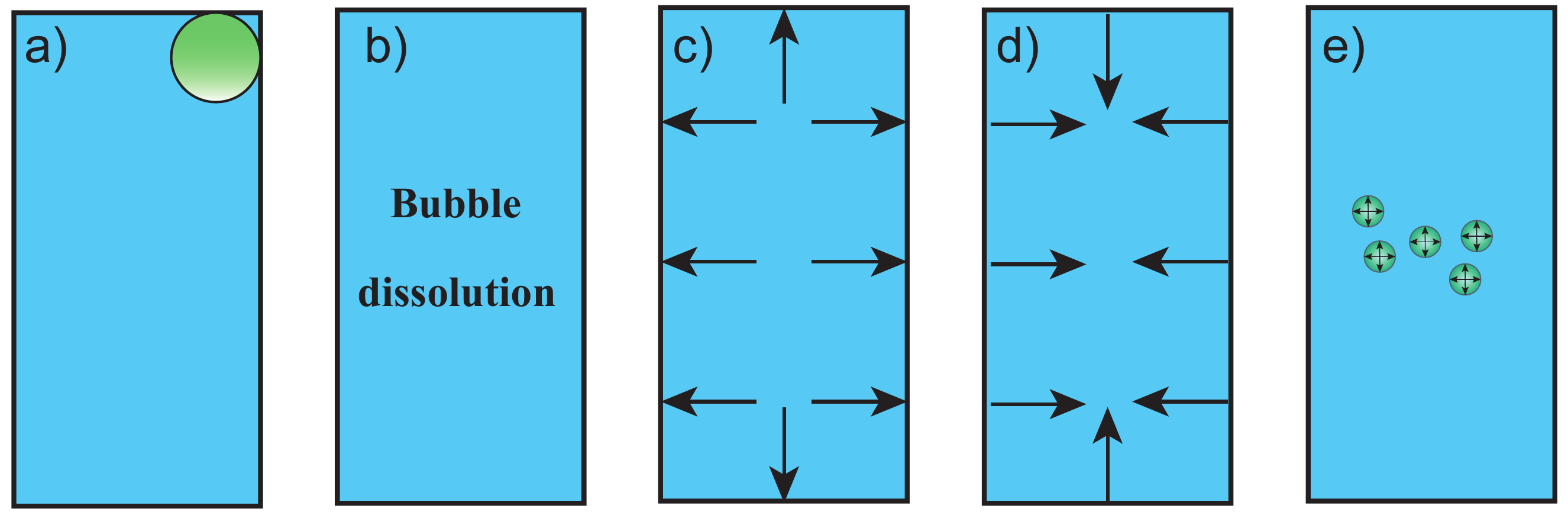}
\caption{\footnotesize (Color online) Schematic representation of the formation of vapor cavities/bubbles (green spheres) within a dominant phase of water (blue background). In panel a), a container of water is sealed containing a small air bubble. Upon increasing $T$ from a) to b), water expands and the bubble dissolves into the liquid; c) upon further heating, the walls of the vessel (black arrows) continue to increase; d) the system is cooled slowly in order to reach a metastable state, which in turn gives rise to the detachment of the liquid from the vessel's walls and, as a consequence, e) cavitation of small air bubbles occurs. Panels a), b), and c) represent a positive pressure regime, while d) and e) represent a negative one. Figure adapted from Ref.\,\cite{imre}.}
\label{Fig-3}
\end{figure}
The nucleation rate $k$ is given by an Arhenius-type behavior $k = k_0\exp{(-E_b/k_B T)}$ \cite{cavitation}, where $k_0$ is a pre-factor, making cavitation favored when $E_b$ is comparable to the thermal energy \cite{maris}.
In this context, the outward pressure the vapor bubbles exert on the metastable liquid can be interpreted as a negative pressure, whereas the liquid applies an inward (positive) pressure on the bubbles, see Fig.\,\ref{Fig-3}. Upon increasing $x$, a liquid-gas phase transition-like behavior occurs. The vapor phase (phase B in Fig.\,\ref{Fig-2}) dominates the system, which might coexist with liquid droplets. The latter situation occurs outside of the nucleation regime and thus is not covered by nucleation theory \cite{sethna}. Instead, a useful quantity for analyzing phase transitions, metastability, and critical phenomena, is the Gr\"uneisen ratio $\Gamma = \alpha_p/c_p \simeq (\partial S/\partial p)_T/(\partial S/\partial T)_p$, where $\alpha_p$ is the thermal expansivity and $c_p$ the specific heat, both at constant pressure \cite{zhu,mrb}. When a phase transition is induced by varying, for instance, $p$ or $T$, the entropy is dramatically altered and this is reflected in $\Gamma$. It turns out that $\Gamma$ is the singular portion of the so-called effective Gr\"uneisen parameter $\Gamma_{eff}$ \cite{zhu}, which quantifies how $p$ varies with $E$. Thus, using Eq.\,\ref{bindingenergy}, we compute $\Gamma_{eff}$ aiming to connect it with $k$. Now, we make an analysis of the inherent negative pressure in the nucleation process using the effective Grüneisen parameter $\Gamma_{eff}$, defined as \cite{EJP}:
\begin{equation}
\Gamma_{eff}=\frac{v_0\alpha_p}{\kappa_T c_v}=v_0\left (\frac{\partial p }{\partial E}  \right )_v,
\end{equation}\linebreak
where $\kappa_T$ is the isothermal compressibility,
\begin{equation}
\kappa_T = -\frac{1}{v}\left(\frac{\partial v}{\partial p}\right)_T,
\label{kappatexp}
\end{equation}
$c_v$ heat capacity at constant volume, and $v_0$ a reference volume. Due to the fact that both $\Gamma$ and $\Gamma_{eff}$ incorporates various thermodynamic observables that depend on the entropy, both quantities are sensitive on the verge of phase transitions and critical points given the intrinsic entropy accumulation in such a regime \cite{EJP}. We compute $\Gamma_{eff}$ straightforwardly by differentiating Eq.\,\ref{bindingenergy} with respect to $(p_v - p_l)$, which gives:
\begin{equation}
	\Gamma_{eff}=-\frac{3v_0(p_v-p_l)^{3}}{32\pi\sigma^3}.
\label{gammanucleation}
\end{equation}
By analysing Eq.\,\ref{gammanucleation}, it can be inferred that for $p_l < p_v$, which chracterizes the negative pressure regime \cite{sethna}, $\Gamma_{eff} < 0$. On the other hand, for $p_l > p_v$, $\Gamma_{eff} > 0$. Hence, in the context of the nucleation process, $\Gamma_{eff}$ is naturally linked with both negative and positive pressures. By replacing $(p_v - p_l) = -\rho_l \mathscr{L}\Delta T/T_v$ \cite{sethna} into Eq.\,\ref{gammanucleation}, we achieve:
\begin{equation}
\Gamma_{eff} = \frac{3v_0}{32\pi\sigma^3}\left[\rho_l\mathscr{L}\frac{(T_v - T)}{T_v}\right]^3,
\end{equation}
where $\rho_l$ is the density of the liquid, $\mathscr{L}$ the latent heat, $\Delta T = (T_v - T)$, and $T_v$ the vapor temperature. It becomes then clear that $\Gamma_{eff} \propto [(T_v - T)/T_v]^3$ for nucleation processes, being zero for $T = T_v$ and enhanced upon decreasing $T$ below $T_v$. Note that for $T = T_v$ or $p_l = p_v$, $E_b \rightarrow \infty$ meaning that nucleation is prevented exactly in the change of regime from stable to metastable       \cite{sethna}. Such a feature is reinforced by the fact that when $E_b \rightarrow \infty$ $\Rightarrow$ $k \rightarrow 0$. For small supercoolings, namely for temperatures close to $T_v$, the nucleation rate is tiny and it grows as $T$ is decreased below $T_v$ \cite{sethna}. Hence, it becomes evident that $\Gamma_{eff}$ applied to nucleation theory (Eq.\,\ref{gammanucleation}) naturally incorporates the concept of negative pressure. Also, $\Gamma_{eff}$ is linked with the nucleation rate, since it quantifies the pressure dependence of $E_b$ (Eq.\,\ref{bindingenergy}).


\section{The capillary effect in plants}

\begin{figure}[!htb]
\centering
\includegraphics[width=0.43\columnwidth]{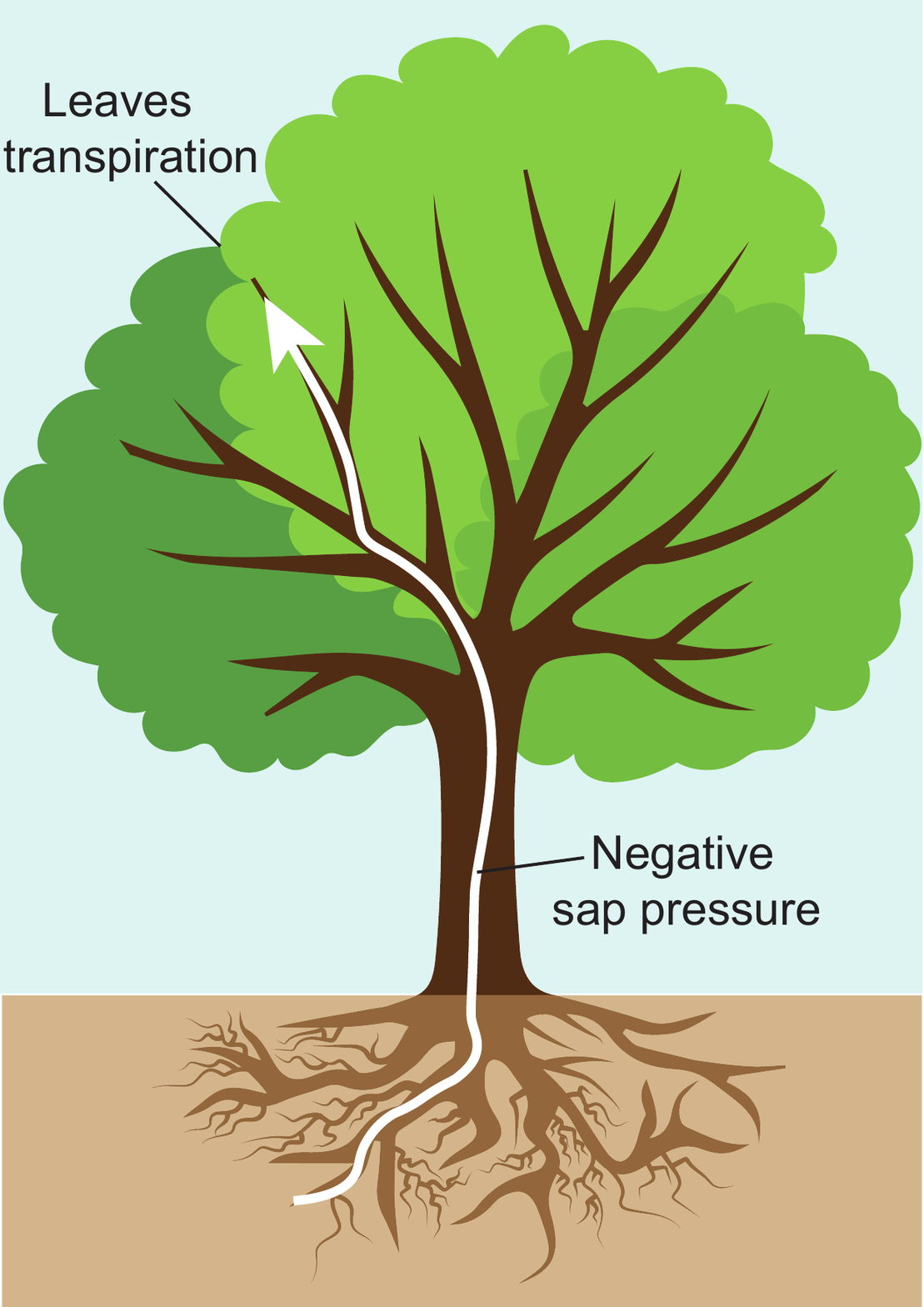}
\caption{\footnotesize Schematic representation of negative sap pressures in plants. Given the water loss through the leaves due to transpiration, a pressure gradient is created, so that a negative sap pressure pulls water from the tree roots \cite{holbrook}, and up through the plant, as indicated by the white arrow.}
\label{Fig-4}
\end{figure}

In this section we review the role negative pressure plays in the capillary effect in plants. Plants use their vascular system to extract water from the soil; this involves the so-called negative sap pressure \cite{hammel}, see Fig.\,\ref{Fig-4}.
The evaporation of water from the leaves creates a negative pressure and, because of the cohesion forces between water molecules, this negative pressure pulls water from the soil, creating a continuous hydration cycle that is essential for plant survival \cite{stroock,nobel}. Astonishingly, negative pressures in plants can be as large as $-$80\,bar for some desert species \cite{holbrook}. Due to the competition between water molecule adhesion to the plant capillary and the cohesive forces among molecules, water is in a metastable phase under these conditions \cite{nobel}. Although cavitation occurs in the xylem, the formed gas bubbles are dissolved, otherwise an embolism would be formed preventing the capillary effect \cite{nobel}. \newpage

\section{Tensile stress as negative pressure}\label{tensilestress}
We now shift our focus from liquids to solids. In some cases, negative pressures emerge due to tensile stresses. In this section, we discuss two particular situations where tensile stress plays a crucial role: the deuteration of the methyl-end groups of molecular conductors and negative thermal expansion.
\subsection{Molecular metals and negative chemical pressure}

In molecular conductors of the (TMTTF)$_2$X family, where TMTTF is the base molecule tetramethyltetrathiafulvalene and $X$ is a monovalent counter-anion\footnote{The term counter-anion here is used in analogy with, for instance, NaCl where Na$^+$ is the cation and Cl$^-$ the counter-anion. Hence, TMTTF$^+$ and PF$_6$$^-$ or SbF$_6$$^-$ play the respective roles of Na$^+$ and Cl$^-$.} such as PF$_6$ or SbF$_6$, negative pressure is associated with the so-called chemical pressure \cite{prbrioclaro}. In the particular case of the TMTTF-based charge-transfer salts, replacing the counter-anion changes the molecular mass, modifies the overlap of orbitals, and alters the position of the system in the phase diagram \cite{reviewmariano}. This gives rise to chemical pressure, which in turn dictates the physical properties of the system. Such molecular metals have a triclinic crystalline structure in which the counter-anions are trapped in cavities delimited by the methyl-end groups of the TMTTF molecules \cite{prbrioclaro} (see Fig.\,\ref{Fig-1} c). It turns out that deuteration (i.e., the process of chemically replacing hydrogen atoms by deuterium) of such methyl-end groups increases the volume of the cavities lodging the counter-anions because the C-D bond length is smaller than that of C-H by $\sim$0.01\,\AA\,\cite{pouget}. This in turn acts as a negative pressure on the system increasing the charge-ordering transition temperature \cite{pouget}. It is clear that the negative pressure in these molecular conductors is associated with tensile stress. The shortening of the C-D bond length emulates the detachment of the counter-anion from the ``vessel walls'', i.e., the cavity delimited by the methyl-end groups. As a consequence, the configurational entropy is increased because of the enhanced mobility of the counter-anions.

\subsection{Negative thermal expansion, negative pressure, and metastability}

\noindent The thermal expansion coefficient at constant pressure is defined as
\begin{equation}
\alpha_p =\frac{1}{v}\left(\frac{\partial v}{\partial T}\right)_p.
\label{thermalexp}
\end{equation}
Usually, upon cooling a system its volume decreases so $\alpha_p > 0$. However, in rare cases, negative thermal expansion takes place, i.e., the system expands upon cooling so that $\alpha_p < 0$. This occurs, for instance, in water (for certain temperature ranges, or when supercooled) \cite{allan}, at the Mott transition \cite{negativealphamott}, and in other specialized materials such as ZrW$_2$O$_8$ \cite{allan}, Y$_2$Mo$_3$O$_{12}$ \cite{lind}, Ag$_3$[Co(CN)$_6$] \cite{pnas}, and black phosphorus \cite{reily}.

A negative $\alpha_p$ results from the metastable character of the system \cite{allan}. The connection between negative $p$ and negative $\alpha_p$ follows from the cyclic relation:
\begin{equation}
\left(\frac{\partial v}{\partial p}\right)_T \left(\frac{\partial p}{\partial T}\right)_v\left(\frac{\partial T}{\partial v}\right)_p = -1.
\end{equation}
With the help of the reciprocal relation $(\partial T/\partial v)_p = [(\partial v/\partial T)_p]^{-1}$ along with the definition of $\kappa_T$ and $\alpha_p$ from Eqs.\,\ref{kappatexp} and \ref{thermalexp}, this equation can be rearranged to:
\begin{equation}
\left(\frac{\partial p}{\partial T}\right)_v = \frac{\alpha_p}{\kappa_T},
\label{keyexpression}
\end{equation}
which is the thermal pressure coefficient \cite{imre}.

At low temperatures, some crystalline materials may be under internal stress that eventually causes the system to expand upon cooling \cite{allan}. In other words, when $T$ is reduced, structural metastability gives rise to an internal outward pressure, i.e., a negative pressure/tensile stress, making the system expand. The particular physical mechanism associated with the appearance of structural metastability (and tensile stresses) depends on the system. For an example involving strongly correlated electronic systems, we note that the electron localization at the Mott transition can be seen as analogous to the detachment of liquid from the vessel walls discussed previously. It turns out that, upon localization, itinerant electrons no longer contribute to the solid's cohesion, therefore the Fermi pressure is reduced and the solid expands \cite{negativealphamott}. It is then clear that the negative thermal expansion is driven by a negative pressure. Hence, one can straightforwardly see that the physical origin of a negative $\alpha_p$ is connected with the emergence of metastability, namely negative pressure/tensile stress.

\section{Conclusions}
We have discussed the concept of negative pressure in various scenarios, as well as the interdisciplinary character associated with metastability and cavitation. We also explored the connection between negative thermal expansion and negative pressures. For the cavitation process in the context of nucleation theory we have employed $\Gamma_{eff}$ and demonstrated that it is dramatically enhanced when the energy scale associated with a vapor bubble vanishes. Our work provides perspective by highlighting the shared physical aspects associated with the appearance of negative pressures in distinct fields of research. We hope that the present work benefits undergraduate students by introducing the concept of negative pressure in a way, not usually covered in classical textbooks.

\section{Authors Declarations}
\subsection*{Conflict of interests}
The authors have no conflicts to disclose.

\section*{Acknowledgements}
MdeS acknowledges financial support from the S\~ao Paulo Research Foundation - Fapesp (Grants No.\,2011/22050-4, 2017/07845-7, and 2019/24696-0), National Council of Technological and Scientific Development - CNPq (Grant No.\,303772/2023-9). ACS acknowledges CNPq (Grant No.\,08695/2021-6). LSR acknowledges the Icelandic Research Fund (Rannis) (Grant No.\,163082-051). LS acknowledges IGCE for the post-doc fellowship.








\end{document}